\documentclass[titlepage,12pt]{article}

\begin{document}

\begin{titlepage}

\vskip .6in

\begin{center}
{\Large {\bf   Higher Order Spin-dependent Terms in D0-brane Scattering
 from the Matrix Model  }}

\end{center}

\normalsize
\vskip .6in

\begin{center}

I. N. McArthur
\par \vskip .1in \noindent
 {\it Department of Physics, The University of Western Australia}\\
{\it Nedlands, W.A. 6907. Australia }

\end{center}
\vskip 3cm

\begin{center}
{\large {\bf Abstract}}\\
\end{center}

The potential describing long-range interactions
 between D0-branes contains spin-dependent terms. In
 the matrix model, these should be  reproduced by the
one-loop effective action computed in the presence of a nontrivial
fermionic background $\psi.$ The $\frac{v^{3} \psi^{2}}{r^{8}}$ term
in the effective action has been computed by Kraus and shown to
correspond to a spin-orbit interaction between D0-branes, and the
 $\frac{ \psi^{8}}{r^{11}}$ term in the static potential has been obtained
 by Barrio et al. In this
paper, the $\frac{v^{2} \psi^{4}}{r^{9}}$ term is computing in the
matrix model and compared  with the corresponding results of
Morales et al obtained using string theoretic methods. The technique employed
is adapted to the underlying supersymmetry of the matrix model, and
should be useful in the calculation of  spin-dependent effects in more
general Dp-brane scatterings.

\vspace{9cm}
\noindent

\end{titlepage}

\section{Introduction}
Recent developments in superstring duality owe much to the
recognition of the role of Dirichlet-branes as BPS states which act
as  sources for Ramond-Ramond fields \cite{Polch1}. The long-distance
interactions of D-branes via the exchange of a  closed string have been
studied, and the static potential vanishes \cite{Polch1}, as is
characteristic for BPS states. The leading term in the long range
velocity-dependent potential for a pair of slowly moving  D0-branes behaves as
$\frac{v^4}{r^7},$ where $v$ is the relative speed and $r$ is the
separation \cite{Bachas,Lif}.  This result can be reproduced from the
low energy effective theory  describing the dynamics of D0-branes in
terms the  states of  open strings
which end on the D0-branes. The low energy effective theory is  the
dimensional
reduction of
 10D supersymmetric Yang Mills theory to 1+0
dimensions \cite{Witten,Dan,Kabat,Douglas}, and the potential follows from
 the one loop effective action computed in the presence of a nontrivial
background containing information about the relative motion of the D0-branes.
These results  also provide important tests of the conjecture by
Banks, Fischler, Shenker and Susskind (BFSS) that the dynamical degrees of
freedom of eleven-dimensional M-theory in the infinite momentum frame
are D0-branes \cite{BFSS}. Low energy D0-brane scattering amplitudes
computed using the dimensionally reduced super Yang-Mills theory (a
quantum mechanical
matrix model) reproduce tree level scattering amplitudes for
eleven-dimensional supergravitons.

As BPS states, D0-branes belong to a shortened
supersymmetry multiplet, and the leading term in the velocity-dependent
potential
is the same for all spin states in the supermultiplet. Spin dependent terms
in the long range potential were considered by Harvey \cite{Harvey},
who pointed out that they
 should be reproduced in the matrix model by   calculation of the one loop
effective
action in the presence of a  fermionic
background $\psi.$ Comparison of these results with the spin dependence of
supergraviton scattering in eleven dimensions will be important tests of
the BFSS conjecture. On dimensional grounds, the spin dependence of the
potential was argued in \cite{Harvey} to be coded in an effective
action of the form
\begin{equation}
 \frac{v^4}{r^7} +  \frac{v^3 \psi^2}{r^8} + \frac{v^2 \psi^4}{r^9} +
 \frac{v \psi^6}{r^{10}} + \frac{\psi^8}{r^{11}}.
 \label{spin}
 \end{equation}
The order $\psi^2$ term in this expansion has been computed in the
matrix model by Kraus \cite{Kraus}, who showed that it reproduces the
spin-orbit interaction for a D0-brane probe moving in the
linearized metric of a spinning D0-brane. This interaction has also
computed using string theoretic methods by Morales et al
\cite{Morales1,Morales2}.

The  matrix model calculation carried out in \cite{Kraus} was done by
 treating the
fermionic part of the  background as a perturbation about a purely bosonic
background. The order $\psi^8$ term in the static potential has been
obtained by similar means in \cite{Barrio}.
Here, we carry out the computation of the order $\psi^4$ term in the
 matrix model effective action (\ref{spin}) in a manner which is better
adapted
to the underlying supersymmetry.
This is achieved by recognizing that the one loop
effective action in the presence of background fields is related to
the superdeterminant of the operator which appears in the part of the
action quadratic in the quantum fields. In the case where there are no
fermionic background fields, this superdeterminant factorizes into a
number of ordinary determinants, which are easily computed using
Schwinger's proper time formalism \cite{Douglas}. In the presence of a
nontrivial
fermionic background, the superdeterminant no longer factorizes. It
 can still conveniently be  computed using the Schwinger
proper time formalism, but one has to work a little harder than in the
case of a purely bosonic background.

The outline of the paper is as  follows. In \S2, the one-loop
effective action  for the matrix model in the presence of a fermionic
background is formulated in terms of a superdeterminant. The approach
to be taken to evaluate  the one-loop effective action is  illustrated
in \S3
by reconsideration of the well known spin-independent terms  in the
potential between two D0-branes.
The extension of this formalism in the case of a superdeterminat is
carried out in \S4, and \S5 gives the form of  the one-loop effective
action to order $\psi^{4}.$ This is compared with string theoretic
 calculations by
Morales et al \cite{Morales2} in \S6. The paper concludes with a
discussion of the relevance of the techniques employed in this paper
to calculations of  more general Dp-brane scattering amplitudes. A
number   of spinor identities and details of computations are
contained in two appendices.

\section{The One-loop Effective Action as a Superdeterminant}
The matrix model provides a description of the low energy dynamics of
D0-branes in terms of a quantum mechanical model obtained by the
dimensionional reduction of 10D supersymmetric Yang-Mills theory to
$1+0$  dimensions. In particular, for gauge group SU(2) and an
appropriate choice of background fields, the one-loop effective
action for this theory yields the long-range potential between a pair
of D0-branes. In the presence of a fermionic background, this
one-loop effective action is a superdeterminant, which we manipulate
into a  convenient form in this section.

The ten-dimensional supersymmetric Yang-Mills action  (using the
conventions of \cite{GSW}) is
$$ S = \int d^{10}x \, \mbox{ Tr} \, ( - \frac{1}{4}\,F_{\mu \nu}F^{\mu
\nu} + \frac{i}{2} \bar{\psi} \Gamma^{\mu}D_{\mu} \psi ),$$
where $A_{\mu}$ and $\psi$ take values in the Lie algebra of $SU(2),$
and $\psi$ is a  sixteen-dimensional Majorana-Weyl spinor (spinor
conventions are detailed in Appendix A).
 In computing the effective action, the
  fields are decomposed as a sum of background  and quantum
 pieces, $A_{\mu} = A^{(b)}_{\mu} + A^{(q)}_{\mu},$
$\psi = \psi^{(b)} + \psi^{(q)}.$   Since the terms (\ref{spin}) of
interest contain no derivatives of the background spinor fields,
they will be chosen to be constant. Only the piece of the action
quadratic in the quantum fields is relevant in computing the one-loop
effective action.
Choosing  background field gauge, and including
quantum ghost
fields $c^{(q)}$ and $c^{*(q)}$ but no ghost background, this can be
written (after rotation to a Euclidean metric)
\begin{eqnarray}
 S_{quad} & = & \int d^{10}x \,   \left( \frac12 \,A^{(q)}_{\mu}
\,(D^{(b)}_{\rho} D^{(b)}_{\rho} \,  \delta_{\mu \nu} - 2 i  F^{(b)}_{\mu \nu})
 A^{(q)}_{\nu} + \frac{i}{2} \bar{\psi}^{(q)}
\Gamma_{\rho} D^{(b)}_{\rho} \psi^{(q)} \right. \nonumber \\
&- &  \left.   \frac{1}{2} A^{(q)}_{\mu} \bar{\psi}^{(b)}
\Gamma_{\mu} \psi^{(q)}
 - \frac{1}{2} \bar{\psi}^{(q)}
\Gamma_{\nu}  \psi^{(b)} A^{(q)}_{\nu} + c^{*(q)} \,D^{(b)}_{\mu}
D^{(b)}_{\mu}\,c^{(q)} \right).
\label{1}
\end{eqnarray}
In (\ref{1}), there is no trace because the background fields are taken
to be matrices in  the adjoint representation of
SU(2), and are contracted with quantum fields which in turn are
 vectors in the adjoint representation.  The covariant
derivatives with respect to background fields are defined by
 $ D^{(b)}_{\mu} \psi^{(q)} = \partial_{\mu} \psi^{(q)} - i
A^{(b)}_{\mu} \psi^{(q)},$ and $F^{(b)}_{\mu \nu} = \partial_{\mu}
A^{(b)}_{\nu} -  \partial_{\nu}
A^{(b)}_{\mu} - i [ A^{(b)}_{\mu}, A^{(b)}_{\nu} ].$
Upon reduction to (1+0) dimensions with coordinate $\tau,$ this  piece
of the action becomes (with $i = 1, \cdots , 9$):
$$
S_{quad} = \int d\tau \,\left( \frac{1}{2}\Phi^{*} \Delta_1 \Phi + c^{*(q)}
 \,(D^{(b)}_{\tau} D^{(b)}_{\tau} -
  A^{(b)}_{i} A^{(b)}_{i}) \,c^{(q)} \right)
$$
where $$\Phi^{*} = ( A^{(q)}_{\mu}, \bar{\psi}^{(q)}) ,
\, \, \, \, \, \Phi = \left( \begin{array}{l} A^{(q)}_{\nu} \\ \psi^{(q)}
\end{array} \right),
$$
and $\Delta_1$ is the operator
$$ \left( \begin{array}{ll}    (D^{(b)}_{\tau}D^{(b)}_{\tau} - A^{(b)}_{i}
A^{(b)}_{i})\delta_{\mu \nu} - 2 i  F^{(b)}_{\mu \nu} &
\,\,\,\,\,\,\,\,\,\, - \bar{\psi}^{(b)}
\Gamma_{\mu} \\ -  \Gamma_{\nu}  \psi^{(b)} & \,\,\,\,\,\,\,\,\,\,
i (\Gamma_0 D^{(b)}_{\tau} - i \Gamma_i A^{(b)}_{i}) \end{array}
 \right).  $$

Integrating out the quantum fields, the one-loop partition function is
\begin{eqnarray}
Z_{1} & = & \int [dA^{(q)}] \, [d\psi^{(q)}] \, [dc^{*(q)}] \,
[dc^{(q)}] \, \exp (- S_{quad})\nonumber \\
& = &\frac{ \det( D^{(b)\,\, 2}_{\tau} - A^{(b)}_{i} A^{(b)}_{i})}{(\mbox{
sdet}\,
\Delta_1)^{1/2}},
\label{Z1}
\end{eqnarray}
where ``det'' and  ``sdet''  denote the functional determinant and functional
superdeterminant respectively. Since it is proposed to compute these
from the functional trace and supertrace  of the heat
kernels associated with the relevant operators, it is necessary to
convert the operator appearing in the superdeterminant into one which
is of Laplace type. We make use of the definition
$$ \mbox{ sdet} \left( \begin{array}{cc} A & \chi \\ \Sigma & B
\end{array} \right) = \frac{\det ( A - \chi \, B^{-1} \,\Sigma )}{\det B}$$
 to write
$$
 \mbox{ sdet}\, \Delta_1  =  (\det i\Gamma.D)^{-1} \, \det
 (  D^{ 2}\delta_{\mu \nu} - 2 i  F_{\mu \nu}  -
 \bar{\psi}
\Gamma_{\mu}
\frac{1}{i(\Gamma .D)}\Gamma_{\nu}
\psi) ,
$$
where the superscript ``(b)'' on
background fields has been dropped since all fields which appear from
now on will be backgrounds fields,  $D^2$ is shorthand for
$D^{2}_{\tau} - A_{i}A_{i}$ and $\Gamma.D$ is shorthand for $(\Gamma_0
D_{\tau} - i \Gamma_i A_i).$
Using $(\Gamma .D)^2 = -
D^{ 2}{\bf 1}_{16} - \frac{i}{2} \Gamma_{\rho  \sigma} F_{\rho
\sigma} \equiv - D^2 {\bf 1}_{16} - \frac{i}{2} \Gamma .F $ (with $
\Gamma_{\rho  \sigma} =
 \frac12 [\Gamma_{\rho},
\Gamma_{\sigma}],$ and $ \{ \Gamma_{\rho},
\Gamma_{\sigma} \} = -2 \delta_{\rho \sigma} {\bf 1}_{16}$ in the
Euclidean metric), this can be rewritten
\begin{eqnarray}
\mbox{ sdet}\, \Delta_1 & = & \left( \det  (D^2 {\bf 1}_{16} +
\frac{i}{2} \Gamma . F) \right)^{-\frac{1}{2}} \,
\, \det \left(   D^{2}\delta_{\mu \nu} - 2 i  F_{\mu \nu}  \right. \nonumber \\
&  - &
\left. \bar{\psi}
\Gamma_{\mu} i(\Gamma .D)
\frac{1}{ (D^{ 2} {\bf 1}_{16} + \frac{i}{2}
\Gamma .F)} \Gamma_{\nu}
\psi \right) \nonumber \\
& = & \left( \det \, (D^2  +
\frac{i}{2} \Gamma .F) \right)^{ \frac{1}{2}} \, \mbox{ sdet} \, \Delta_2,
\label{D1}
\end{eqnarray}
where $\Delta_2$ is the operator
\begin{equation}
 \Delta_2 = \left( \begin{array}{ll}  D^{2}\delta_{\mu \nu}
 - 2i F_{\mu \nu} & \,\,\,\,
\bar{\psi} \Gamma_{\mu} i (\Gamma .D) \\
\Gamma_{\nu} \psi  & \,\,\,\
D^{2}{\bf 1}_{16} + \frac{i}{2}  \Gamma .F
\end{array} \right).
\label{D2}
\end{equation}

Putting the results (\ref{Z1}) and (\ref{D1}) together, the one loop
effective action is
$$ - \ln Z_{1} = - \ln \det D^2 + \frac14 \ln \det \, (D^2 {\bf 1}_{16} +
\frac{i}{2} \Gamma .F) + \frac12 \ln \mbox{sdet} \, \Delta_2.$$
Because all the operators  are of Laplace type,
this can be computed as
\begin{equation}
 - \ln Z_{1} = \int_{0}^{\infty} \frac{ds}{s} \left( \mbox{Tr}\, e^{s
D^2} - \frac14 \mbox{Tr} \, e^{s (D^2 {\bf 1}_{16} +
i \Gamma .F/2)} - \frac12 \mbox{Str}\, e^{s \Delta_2} \right),
\label{effa}
\end{equation}
where Tr and Str denote the functional trace and functional supertrace
respectively, as well as traces over the gauge indices.

The background fields can be decomposed as
$A_{\mu} = A_{\mu}^{+} T_{+} + A_{\mu}^3
T_3  + A_{\mu}^{-}T_{-}$ and $\psi = \psi^{+} T_{+} + \psi^3
T_3 + \psi^{-}T_{-},$ where $(T_+, T_3,T_-)$ are the $3\times3$ matrix
generators of  the
adjoint representation of SU(2). For the case of a scattering of
D0-branes with relative speed $v$ and impact parameter $b,$  the
relevant supersymmetric Yang-Mills background is $ A_1^3 = v \tau, $ $A_2^3
= b,$ and $\psi^3$ constant,
with all other components of $A_{\mu}$ and $\psi$ vanishing
\cite{Witten}. The only
nontrivial components of the Yang-Mills field strength are thus
$F_{01} = - F_{10} = v \, T_3,$ and $\Gamma .F = 2 v \, \Gamma_0
\Gamma_1\,T_3.$ Due to the fact that all of the background fields are
proportional to $T_3,$ the operators in the effective action  (\ref{effa})
decompose into two decoupled pieces with respect to their gauge
indices. The operator acting on the quantum fields
$A_{\mu}^{3(q)},$ $\psi^{3(q)}$ and $c^{3(q)}$ is $\partial_{\tau}^2,$ so
these are massless free fields which decouple \cite{Douglas}; their
contribution to the effective
action (\ref{effa}) is
$$\int_{0}^{\infty} \frac{ds}{s} (1 - \frac14
16 - \frac12 10 + \frac12 16) \mbox{ Tr} \, e^{s \partial_{\tau}^2},$$
which vanishes as a result of a cancellation between bosonic and fermionic
(including ghost) degrees of freedom due to supersymmetry. The other
piece is a $2\times 2$ block with respect to gauge indices,
corresponding to the quantum fields $ (A_{\mu}^{+(q)},
A_{\mu}^{-(q)}),$ $(\psi^{+(q)},\psi^{-(q)})$ and $(c^{+(q)},c^{-(q)}).$
So  the trace over gauge indices in the expression
(\ref{effa}) can be restricted to this $2 \times 2$ block,
 in which the generator $T_3$ accompanying the background fields is
represented by the matrix
\begin{equation} T_{3} = \left( \begin{array}{cc} 1 & 0 \\ 0 & -1
\end{array} \right).
\label{T3}
\end{equation}

With this simplification, determining the one-loop effective
action  reduces to the calculation of the functional traces and
supertraces in the expression (\ref{effa}) in the presence of the
background fields appropriate to the description of D0-brane scattering.

\section{Evaluation of  Functional Traces}
Although the results of computation of the functional traces $\mbox{Tr}\, e^{s
D^2}$  and $ \mbox{Tr}\, e^{s (D^2 {\bf 1}_{16} +
i \Gamma .F/2)}$ are well known \cite{Douglas}, they are rederived here to
illustrate the method that is to be employed to compute the more
complicated functional supertrace $\mbox{Str}\, e^{s \Delta_2}.$
Concentrating on  $ \mbox{Tr}\, e^{s (D^2 {\bf 1}_{16} +
i \Gamma .F/2)},$ this is
$$ \int d \tau \, \lim_{\tau \rightarrow \tau\prime}
\, \mbox{tr} \, e^{s(\partial_{\tau}^2{\bf 1}_2 {\bf 1}_{16} -(v^2 \tau^2 +
b^2)T_3^{\,\,\,2} {\bf 1}_{16} + i v T_3
\Gamma_0 \Gamma_1)} \delta(\tau - \tau'),$$
where ${\bf 1}_2 {\bf 1}_{16}$ denotes the tensor product of the unit
matrices in the two-dimensional representation of the gauge group and
the sixteen dimensional spinor representation, and ``tr'' is the trace
over these representations.
Representing the delta function in the form $\int \frac{dk}{2\pi}
e^{i k (\tau - \tau \prime)},$ this can be written
$$ \int d \tau \int \frac{dk}{2\pi} \mbox{ tr} \left( \,
e^{s( - b^2 {\bf 1}_2 {\bf 1}_{16} +
i v T_3 \Gamma_0 \Gamma_1)} \, e^{s(X^2 - v^2 \tau^2) {\bf 1}_2 {\bf
1}_{16}} \right),$$
where $X = \partial_{\tau} + ik.$
Performing the  traces over the gauge and spinor indices yields
\begin{equation}
\mbox{ Tr}\, e^{s (D^2 {\bf 1}_{16} +
i \Gamma .F/2)} =
32 \int d\tau \, \cosh sv \,\, e^{-s b^2} \, K_0(s),
\label{Tr2}
\end{equation}
 where
$$ K_0(s) \equiv \int \frac{dk}{2\pi} \, e^{s(X^2 - v^2 \tau^2)}.$$
Similarly,
\begin{equation}
\mbox{ Tr}\, e^{s D^2} = 2 \int d \tau \, e^{-s b^2} \,
K_0(s).
\label{Tr1}
\end{equation}

 Thus one is left to evaluate $K_0(s).$
To achieve this, we employ a method introduced in \cite{MG}, which
has the advantage that it extends naturally to the situation to be
encountered when evaluating the functional supertrace
$\mbox{Str}\, e^{s \Delta_2}$ in (\ref{effa}).  Noting
that
\begin{equation}
 \frac{d K_0(s)}{ds} = K_2(s) - v^2 \tau^2 K_0(s)
 \label{deK}
 \end{equation}
with
$$K_{2}(s) \equiv  \int \frac{dk}{2\pi} \, X^2 \,e^{s(X^2 - v^2
\tau^2)},$$
the aim is to express $K_2(s)$ in terms of $K_0(s)$ so that
the differential equation (\ref{deK}) for $K_0(s)$  can be
solved. To do this, one uses the fact that
$$\int \frac{dk}{2\pi} \, \frac{\partial}{\partial k} \left( X \,e^{s(X^2 -
v^2
\tau^2)}\right) = 0.$$
Performing the derivative gives
\begin{eqnarray*}
 0 & = & i K_0(s) +\int \frac{dk}{2\pi}\, X \int_0^1 du \, e^{us(X^2 - v^2
\tau^2)} \, 2 i s X \,e^{(1-u)s(X^2 - v^2
\tau^2)} \\
& = &i K_0(s) + 2 i \sum_{n=0}^{\infty}\, \frac{s^{n+1}}{(n+1)!}
\int \frac{dk}{2\pi} \,X \,\mbox{ ad}^{(n)} (X^2 + v^2 \tau^2)(X)
\,e^{s(X^2 - v^2
\tau^2)},
\end{eqnarray*}
where $\mbox{ ad}^{(n)}(A)(B)$ stands for the $n$ commutators
$[A,[A, \cdots [A,B] \cdots]].$ The commutators are easily computed
and the series summed to give
\begin{equation}
 0 = K_0(s) + \frac{\sinh 2 s v}{v}\, K_{2}(s)  + \tau (\cosh  2 s v
-1) K_1(s) + (\cosh  2 s v -1) \, K_0(s),
\label{K2}
\end{equation}
where $K_1(s) \equiv  \int \frac{dk}{2\pi} \, X\,e^{s(X^2 - v^2
\tau^2)}.$
The latter is computed in terms of $K_0(s)$ from the identity
$$ 0 =  \int \frac{dk}{2\pi} \, \frac{\partial}{\partial k}\,e^{s(X^2 - v^2
\tau^2)},$$
which yields
$$ K_1(s) = - \tau v\, \frac{(\cosh 2 s v -1)}{\sinh 2 s v} \, K_0(s).$$
Solving (\ref{K2}) for $K_{2}(s)$ and substituting into (\ref{deK})
gives the differential equation
\begin{equation}
 \frac{d \ln K_0(s)}{ds} = -v^2 \tau^2 + v^2 \tau^2 \frac{(\cosh 2 s
v - 1)^2}{\sinh^2 2sv} - v \, \frac{\cosh 2sv}{\sinh 2 s v} ,
\label{deK1}
\end{equation}
which is solved (with the boundary condition that $ K_0(s) =
(4 \pi s)^{- \frac{1}{2}}$ in the limit $v \rightarrow 0$) by
\begin{equation}
K_0(s) = \left( \frac{v}{2 \pi \sinh 2 s v}\right)^{\frac12}\,
 e^{- v \tau^2 \tanh s  v}.
 \label{K0}
 \end{equation}
Substituting this result into (\ref{Tr2}) and (\ref{Tr1}) yields the
standard expressions for the functional traces
 $\mbox{ Tr}\, e^{s (D^2 {\bf 1}_{16} +
i \Gamma .F/2)}$ and $\mbox{ Tr}\, e^{s D^2}.$

\section{Evaluating the Functional Supertrace}
The procedure in the previous section  will be adapted
 to evaluate
$\mbox{Str}\, e^{s \Delta_2}$ in the effective action (\ref{effa}).
Using (\ref{D2}), and replacing the delta function in the
functional supertrace by its Fourier representation,
\begin{equation}
\mbox{Str}\, e^{s \Delta_2} =
\int d\tau  \int \frac{dk}{2 \pi} \, \mbox{ str}\, e^{s \Delta}
\equiv \int d\tau \, \tilde{K}_0(s) ,
\label{D}
\end{equation}
 where
$$ \Delta = \left( \begin{array}{ll} (X^2 - v^2 \tau^2 - b^2) {\bf 1}_{10}
{\bf 1}_{2} - 2i F &  i \bar{\psi} \Gamma (\Gamma_0 X - i \Gamma_1 v \tau T_3
 -  i \Gamma_2 b T_3 )   \\
\Gamma \psi & (X^2 - v^2 \tau^2 - b^2) {\bf 1}_{16} {\bf 1}_2 + \frac{i}{2}
\Gamma .F \end{array} \right)
$$
with $X = \partial_{\tau} + i k.$  The supertrace ``str''
involves an ordinary trace over gauge indices, and a supertrace over
Lorentz indices in the vector and spinor representations
 (which have been suppressed).
We choose to write $\Delta$ in the form
$$ \Delta = \tilde{X}^2  - (v^2 \tau^2 + b^2) {\bf 1} + F.\Lambda +
Y.$$
Here, ${\bf 1}$ is the tensor product of unit matrices in the gauge and
Lorentz
representations,
$$ {\bf 1} = \left( \begin{array}{cc} {\bf 1}_{10} {\bf 1}_2 & 0 \\
0 & {\bf 1}_{16} {\bf 1}_2 \end{array} \right). $$
Also,
$$F.\Lambda = \left( \begin{array}{cc} -2 i F_{\mu \nu} & 0 \\ 0 &
\frac{i}{2} F_{\rho \sigma} \Gamma_{\rho \sigma} \end{array} \right)$$
is the contraction of the Yang-Mills field strength  with a supermatrix
 containing the generators of the
vector and  (Weyl) spinor representation of the Lorentz group in its diagonal
blocks.  The operator $$\tilde{X} = (\partial_{\tau} + i k) {\bf 1} +
N$$ contains  the supermatrix
\begin{equation}
 N = \left( \begin{array}{cc} 0 & \frac{i}{2} \bar{\psi} \Gamma_{\mu}
\Gamma_0 \\ 0 & 0 \end{array} \right).
\label{N}
\end{equation} It can be viewed as a supercovariant derivative
(shifted by $ik).$
Finally, $Y$ is the off-diagonal supermatrix
\begin{equation} Y = \left( \begin{array}{cc} 0 & \bar{\psi} \Gamma_{\mu}
\Gamma_i A_i \\
\Gamma_{\nu} \psi & 0 \end{array} \right)
\label{Y}
\end{equation}
with $\Gamma_i A_i = \Gamma_1 v \tau T_3 + \Gamma_2 b T_3$
(note that we have used the fact that $N^2 = 0,$ which eliminates a
potential contribution to $Y).$

As in the case of the functional trace evaluated at the start of this
section, the quantity  $\tilde{K}_0(s) = \int \frac{dk}{2 \pi} \,  e^{s
\Delta}$ in (\ref{D}) is computed  using the  fact that it
satisfies the  differential equation
\begin{equation}
 \frac{d \tilde{K}_0(s)}{ds} = \tilde{K}_2(s) +  ( - v^2 \tau^2 -
b^2 + F.\Lambda  + Y) \, \tilde{K}_0(s),
\label{eq1}
\end{equation}
where $\tilde{K}_n(s) = \int \frac{dk}{2 \pi} \, \tilde{X}^n\, e^{s
\Delta}.$  As before, the requirement is to express $\tilde{K}_2(s)$
in terms of $\tilde{K}_0(s)$ to allow this differential equation to
be solved. $\tilde{K}_2(s)$ is computed with the aid of the identity
\begin{eqnarray}
0 & = & \int \frac{dk}{2 \pi} \, \frac{\partial}{\partial k} \,
\tilde{X} \,
e^{s \Delta} \nonumber \\
& = & i \tilde{K}_0(s) + 2 i \, \sum_{n=0}^{\infty} \frac{s^{n+1}}{(n+1)!} \,
\int \frac{dk}{2 \pi} \, \tilde{X} \, ad^{(n)}(\Delta)(\tilde{X}) \,
e^{s \Delta}.
\label{eq3}
\end{eqnarray}
Recalling that  we  are interested in the term in (\ref{spin})
 of fourth order in the
 background fermions
$\psi,$ it is easily established that the commutators have the following
structure to this order:
\begin{eqnarray}
ad^{(2n)}(\Delta)(\tilde{X}) & = & 2^{2n} v^{2n}  \tilde{X} +
N_{2n} \tilde{X} + Y_{2n}\nonumber \\
ad^{(2n+1)}(\Delta)(\tilde{X}) & = & 2^{2n+1} v^{2n+2} \tau {\bf 1} +
N_{2n+1} \tilde{X} + Y_{2n+1},
\label{commutators}
\end{eqnarray}
with
\begin{equation}
 N_1 = 0,  \,\,\, N_n =
2[\tilde{X},Y_{n-1}] + [F.\Lambda + Y, N_{n-1}] \, \, \, (n \geq 2)
\label{defN}
\end{equation}
and
\begin{eqnarray}
 Y_1 &=& [F.\Lambda + Y, \tilde{X}], \, \, \, \, Y_{2} =  [F.\Lambda +
 Y, Y_{1}], \nonumber \\
 Y_{2n+1} & = & 2 \tau v^2 N_{2n} + 2^{2n} v^{2n} Y_1 + N_{2n} Y_1 +
  [F.\Lambda + Y,
Y_{2n}], \nonumber\\
 Y_{2n+2} & = & 2 \tau v^2 N_{2n+1} + N_{2n+1} Y_1 + [F.\Lambda + Y,
Y_{2n+1}] \, \,\, (n \geq 1).
\label{defY}
\end{eqnarray}
The general form of the supermatrices $N_n$ and $Y_n$ in powers of
$\psi,$  $\tau$ and $b$ can be established inductively to be
\begin{equation}
 N_n = \left( \begin{array}{cc} O(\psi^4) & O(\psi^3) \\
0 & O(\psi^4) \end{array} \right), \,\, Y_n = \left(
\begin{array}{cc} O(\psi^2) + A_{i} O(\psi^4) & O(\psi) + A_{i} O(\psi^3) \\
O(\psi^3) &  O(\psi^2) + A_{i} O(\psi^4) \end{array} \right).
\label{general}
\end{equation}
As a result, $[\tilde{X}, [\tilde{X}, Y_n ]] = 0$ and  $[\tilde{X},
N_n] = 0$ to order $\psi^4,$ eliminating some potential additional
contributions to (\ref{commutators}), (\ref{defN}) and (\ref{defN}). The
explicit form of the
matrices $N_n$ and $Y_n$ is required only for $N_{1}, N_{2}, N_{3}$ and
$Y_{1},$ and
these are  presented in Appendix B.

Using these results, (\ref{eq3}) then yields
\begin{eqnarray}
0 &=& \left( \frac{\sinh 2 s v}{v} + 2 N(s) \right) \tilde{K}_2(s) \nonumber \\
& + & \left( \tau (\cosh 2 s v -1) + 2 [\tilde{X}, N(s)] + 2 Y(s)
\right) \, \tilde{K}_1(s) \nonumber \\
& + & \left( \cosh 2 s v + 2 [\tilde{X}, Y(s)] \right) \tilde{K}_0(s),
\label{eq6}
\end{eqnarray}
where $$N(s) = \sum_{n=2}^{\infty} \frac{s^{n+1}}{(n+1)!} N_n,
\,\,\,\, Y(s) = \sum_{n=1}^{\infty} \frac{s^{n+1}}{(n+1)!} Y_n. $$
In order to solve for $\tilde{K}_2(s)$ in terms of $\tilde{K}_0(s),$
it is necessary to have an expression for $\tilde{K}_1(s)$ in terms of
$\tilde{K}_0(s).$ This again
follows from the identity
$$ 0 = \int \frac{dk}{2 \pi} \frac{\partial}{\partial k} e^{s
\Delta},$$
which yields
$$ \tilde{K}_1(s) = - \left( \frac{\sinh 2 s v}{v} + 2 N(s)
\right)^{-1} \, \left( \tau (\cosh 2sv -1) + 2 Y(s) \right)
\tilde{K}_0(s).$$
Substituting into (\ref{eq6}),
\begin{eqnarray*}
 \tilde{K}_2(s) &=& v^2 \tau^2\frac{(\cosh 2sv -1)^2}{\sinh^2 2sv}
\left(1 + \frac{2 v}{\sinh 2 s v} N(s) \right)^{-1}.\\
& & \left( 1  +  \frac{2}{\tau (\cosh 2 s v -1)}\, ([\tilde{X}, N(s)] + 2
Y(s)) \right) .\\
& & \left( 1
 +  \frac{2 v}{\sinh 2 s v} N(s) \right)^{-1}\, \left(1 +
\frac{2}{\tau (\cosh 2 s v -1)}Y(s) \right)\, \tilde{K}_0(s) \\
& - & v\, \frac{\cosh 2 s v}{\sinh 2 s v} \left(1 + \frac{2 v}{\sinh 2
s v}\, N(s)
\right)^{-1}.
\\ & &\left( 1
 +   \frac{2}{\cosh 2 s v}\, [ \tilde{X}, Y(s)] \right) \,
\tilde{K}_0(s).
\end{eqnarray*}
This can be simplified considerably using the fact that to order
$\psi^4,$
$$N(s)^2 = 0, \,\,
[\tilde{X}, N(s)] = 0, \,\, N(s) Y(s) = 0, \,\, \mbox{ and} \, \, \,
N(s) [\tilde{X}, Y(s)] = 0,$$
as is easily established from (\ref{general}).
Substituting the simplified expression into (\ref{eq1}) results in
the following differential equation for $\tilde{K}_0(s)$:
\begin{eqnarray*}
\frac{d \ln \tilde{K}_0(s)}{ds} & =& -(v^2 \tau^2 + b^2) {\bf 1} +
F.\Lambda + Y
\nonumber \\
&+ & \tau^2 v^2 \frac{(\cosh 2 s v -1)^2}{\sinh^2 2 sv} \, \left( 1 -
\frac{4v}{\sinh 2 s v} N(s) \right. \nonumber \\
& + & \left. \frac{4}{\tau (\cosh 2 s v-1)} Y(s) +
\frac{4}{\tau^2 (\cosh 2 s v -1)^2} Y(s)^2 \right) \nonumber \\
& - & v \frac{\cosh 2 s v}{\sinh 2 s v} \left( 1 - \frac{2v}{\sinh 2 s
v} N(s) + \frac{2}{\cosh 2 s v} [\tilde{X}, Y(s)] \right).
\end{eqnarray*}
If $\tilde{K}_0(s)$ is factored in the form $\tilde{K}_0(s) = e^{-s
b^2}\, K_0(s) \,
M(s),$  then using the property
(\ref{deK1}), it follows that
\begin{eqnarray}
\frac{d \ln M(s)}{ds} &=& (F.\Lambda + Y) + 4 \tau v^2 \frac{(\cosh 2
s v -1)}{\sinh^2 2 s v} Y(s) \nonumber \\
& + & \frac{4 v^2}{\sinh^2 2 s v} Y(s)^2  - \frac{2v}{\sinh 2 s v}
[\tilde{X}, Y(s)] \nonumber \\
&+& \left( - 4 \tau^2 v^3 \frac{(\cosh 2 s v -1)^2}{\sinh^3 2 s v} + 2
v^2 \frac{\cosh 2 s v}{\sinh^2 2 s v} \right) N(s).
\label{deM}
\end{eqnarray}

Thus, to order $\psi^{4},$ the functional supertrace appearing in the
effective action
(\ref{effa}) has the following expression:
$$ \mbox{Str}\, e^{s \Delta_2} = \int d\tau \, e^{-s b^2} \, K_0(s) \,
\mbox{str} \, M(s),$$
where $K_{0}(s)$ is given in (\ref{K0})  and, to order $\psi^{4},$
 $M(s)$  is the  solution
to the differential equation (\ref{deM}).

\section{The One-loop Effective Action to Order $\psi^{4}$}
Using the above result together with (\ref{effa}), (\ref{Tr2})
and (\ref{Tr1}),  the one-loop effective action for the matrix model
in the presence of background fields relevant to the description of
D0-brane scattering is
$$ -\ln Z_1 = \int d\tau \int_0^{\infty} \frac{ds}{s} \, e^{-s b^2}
\left( 2  - 8 \cosh sv - \frac{1}{2} \mbox{str} M(s) \right) \,
K_0(s).$$
In the absence of a fermionic background, all terms  except the first  on
the right hand side of  (\ref{deM}) vanish, so
that $M(s) = e^{s F.\Lambda}.$ Thus
 $\mbox{ str} M(s) = (4 \cosh 2 s v + 16) - 32 \cosh s v,$
where the $+16$ is the contribution from the piece of the identity
matrix ${\bf 1}_{10} {\bf 1}_2$ which is not in the $2 \times 2$
block in which $F_{\mu \nu}$ is nonvanishing. With $K_0(s)$ as in
 (\ref{K0}), this reproduces the well known results of \cite{Douglas} for
the effective action.  The leading term  in the expansion (\ref{spin})
of the long-range potential between two D0-branes comes from the
order $s^4$ term in the expansion of $\cosh sv$ and $ \cosh 2 s v,$ the
the order $s^0$ terms vanishing because of the equality of the number
of bosonic and fermionic degrees of freedom, and the order $s^2$
terms vanishing because of the vanishing of the supertrace of the
mass squared matrix as a result of supersymmetry.

  In the presence of a fermionic
background, the matrix $M(s)$ is more complicated, giving rise to the
expansion of the effective action in the form (\ref{spin}). In particular,
to order
$\psi^4$ in the background fermions,  $M(s)$ is obtained from
equation (\ref{deM}). The supertrace of $M(s)$ is considered in
Appendix B order by order in its expansion in
powers of $s,$ where it is found (using spinor identities established
in Appendix A) that there is a  nonvanishing $\psi^2$ term at order
 $s^5,$  and the leading $\psi^4$ term emerges at order
$s^6.$ As will be explained below, once the leading $\psi^{2}$ and $\psi^{4}$
terms have been found, it is not necessary to consider such terms at
higher order in the power series expansion in $s.$
The $\psi^2$ term is
\begin{eqnarray}
s^5\mbox{str} M_5 & = &   \frac{s^{5}}{120}\, \mbox{tr}
\, \left( 40 i (F)^{3}_{\mu \nu}G_{\nu i
\mu}
 -  10 i (F)^{2}_{\mu \nu} H_{\nu i \mu} \right. \nonumber \\
 & - & \left. 10i v^{2}F_{\mu
\nu}G_{\nu i \mu} + \frac{5i}{2} v^{2} H_{\mu i \mu} \right) A_{i},
\label{M5}
\end{eqnarray}
where the trace is over gauge indices and
$$G_{\mu \nu \rho} = (\bar{\psi} \Gamma_{\mu \nu \rho} \psi), \, \, \,
H_{\mu \nu \rho} = (\bar{\psi} \Gamma_{\mu} \Gamma_{\nu} \tilde{F}
\Gamma_{\rho} \psi ).$$
Substituting $A_{1}= v \tau,$ $A_{2} = b$ yields  the result
$ - i s^5 \, v^3 b\, \mbox{ tr}\,(\bar{\psi}\Gamma_{012}\psi).$
However, it is also possible to evaluate (\ref{M5}) in a SO(9)
invariant manner using $A_{i} = x_{i} T_{3}$ and $F_{0k} = v_{k}T_{3},$
where  $x_{i}$ and $v_{i}$ $(i = 1, \cdots , 9)$
  are the relative coordinate and speed for the two D0-branes.
In this case
 $$s^5\mbox{str} M_5  = -i s^5 \, v^{2} \, v_{i}\, x_{j} \, \mbox{tr} \,
 G_{0ij}.$$
 Recalling the $\psi =
\psi_3 T_3$ with $T_3$ given by (\ref{T3}), this becomes
$$
s^5 \mbox{str}M_5 = - 2i s^5 \, v^2 \, v_{i}\, x_{j}
\,(\bar{\psi}_3\Gamma_{0ij}\psi_3).
$$

The first nonvanishing $\psi^4$ term occurs at order $s^{6}$ in the
power series expansion of $M(s),$ and is
\begin{eqnarray*}
 s^6 \mbox{str}M_6 & = &\frac{s^6}{720} \, \mbox{tr}\, \left( - 12 F_{\mu \nu}
  G_{\nu i \rho}
F_{\rho \sigma} G_{\sigma j \mu} - 24 (F^{2})_{\mu \nu}G_{\nu i \rho}
G_{\rho j \mu} \right. \\
& + &  \left. 6 F_{\mu \nu} G_{\nu i \rho} H_{\rho j \mu} + 6  G_{\mu i \nu}
F_{\nu \rho} H_{\rho j \mu}
 - \frac{3}{4} H_{\mu i \nu} H_{\nu
j \mu} \right) \, A_i A_j .
\end{eqnarray*}
Again, it  is convenient to evaluate this in a SO(9) invariant form.
Using (\ref{GH}) in Appendix A, and then applying the identity
(\ref{ident4}),
\begin{eqnarray*}
 s^6 \mbox{str}M_6 & = & \frac{s^{6}}{15}\, v_{k} v_{l}x_{i}x_{j} \,
\left( - 15 (\bar{\psi}_{3} \Gamma_{0ik} \psi_{3})\,  (\bar{\psi}_{3}
\Gamma_{0jl} \psi_{3}) \right. \\
 & + & \left.2 \delta_{ij} (\bar{\psi}_{3} \Gamma_{0l \mu} \psi_{3})\,
(\bar{\psi}_{3} \Gamma_{0k \mu} \psi_{3})\, - 3 \delta_{il}
(\bar{\psi}_{3} \Gamma_{0 j \mu} \psi_{3})\,(\bar{\psi}_{3}
\Gamma_{0k \mu} \psi_{3}) \right).
\end{eqnarray*}

Collecting the above results, the effective action is
$$ -\ln Z_1 = \int d\tau \int_0^{\infty} \frac{ds}{s} \, e^{-s b^2}
\left(-s^4v^4 - \frac{ s^5}{2}\, \mbox{str}\,M_5 - \frac{s^6}{2}\,
\mbox{str}\, M_6  \right) \,
K_0(s).$$
From (\ref{K0}),
the term of leading order in $v$ in  $K_0(s)$ is
$$K_0(s) \approx (4 \pi s)^{-\frac12} \, e^{-v^2 \tau^2 s}.$$
So, after a rescaling $s \rightarrow \frac{s}{r^2}$ with $r^2 = b^2 +
v^2 \tau^2,$
\begin{eqnarray}
 -\ln Z_1 & =& \frac{1}{\sqrt{4 \pi}} \,
\int d\tau \int_0^{\infty} ds \, e^{-s}
\left(- s^{5/2}\, \frac{v^4}{r^7} - \frac{s^{7/2}}{2 r^9}
\, \mbox{str} \, M_5 -
\frac{s^{9/2}}{2r^{11}}\, \mbox{str} \, M_6 \right) \nonumber \\
& = & \int d\tau \, \left( - \frac{15}{16}\, \frac{v^4}{r^7} +
\frac{105i}{32}\, \frac{v^2 v_{i}x_{j}}{r^9}\, (\bar{\psi}_3
\Gamma_{0ij} \psi_3)
\right. \nonumber \\
 &+&  \frac{63}{128}\, \frac{1}{r^{11}}\,v_{k}v_{l}x_{i} x_{j}
 \left( 15 (\bar{\psi}_{3} \Gamma_{0ik} \psi_{3})\,  (\bar{\psi}_{3}
\Gamma_{0jl} \psi_{3}) \right. \nonumber \\
&- & 2  \left. \left.\delta_{ij} (\bar{\psi}_{3} \Gamma_{0l \mu} \psi_{3})\,
(\bar{\psi}_{3} \Gamma_{0k \mu} \psi_{3})\, + 3 \delta_{il}
(\bar{\psi}_{3} \Gamma_{0 j \mu} \psi_{3})\,(\bar{\psi}_{3}
\Gamma_{0k \mu} \psi_{3}) \right) \frac{{}}{{}} \right).
\label{result}
\end{eqnarray}
The  order $\psi^{4}$ terms are of the form
$\frac{v^{2}x^{2}\psi^{4}}{r^{11}},$ and correspond to the
$\frac{v^{2}\psi^{4}}{r^{9}}$ terms in (\ref{spin}).

In principle, it is possible  to obtain contributions of the  form
$\frac{v^{3} \psi^{2}}{r^{8}}$ from the order $s^{6}$ term in the expansion
of $M(s),$ namely through terms with the structure $v^{3}x^{3}\psi^{2}.$
However, in practice, this is not possible.
 The coordinate $x_{i},$ in the form of $A_{i},$ is accompanied by a factor
of at
least $\psi$ in all the
supermatrices used in the computation, the lowest order case being in
the off-diagonal entries in
the supermatrix $Y$ in (\ref{Y}). To get it into the diagonal
to appear in supertraces,  it  picks up at least another factor of
$\psi.$ So a term with a factor of $x^{3}$  must be of at least
order $\psi^{6}.$ This argument shows that once a nonvanishing
$\psi^{2}$ or $\psi^{4}$ term is found in the expansion of
 $M(s)$ in powers of $s,$ higher order terms in this expansion need
 not be considered.

\section{Interpretation of the Effective Action}
In this section, we relate the matrix model effective action (\ref{result})
to the
spin dependence of the long-range potential
 between the moving D0-branes. After rotation back to
the Minkowski metric, the  order $\psi^{2}$
term is the same as that obtained by Kraus \cite{Kraus}. By studying
the action for a D0-brane probe moving in the linearized metric of a
D0-brane  with angular momentum, Kraus showed
that this term correctly  reproduces  the spin-orbit interaction provided
$\frac12 (\bar{\psi}_{3}\Gamma_{0ij}\psi_{3})$ is identified with the
angular momentum $J_{ij}$ of the target D0-brane. This term in the
effective action can be written in the form
$$\int d\tau \,V_{spin-orbit}(r) = -\int d \tau \, \frac{15}{16}\,  v^{2} \,
 J_{ij} \, v^{j}
 \partial_{i} \frac{1}{r^{7}},$$
which exhibits it as a derivative of the 9-dimensional Greeen's
function.

According to Harvey \cite{Harvey}, the  order $\psi^{4}$ terms in
 (\ref{result}) should correspond to
higher order spin-orbit couplings or possibly spin-spin couplings. To
obtain an expression of this form, it is necessary to integrate the
last term in (\ref{result}) by parts with respect to $\tau$ by writing
$$ \frac{(x_{i}v_{i}) v_{j}}{r^{11}} = -\frac19 \,
\frac{\partial}{\partial \tau} \left( \frac{x_{j}}{r^{9}}\right) + \frac19 \,
\frac{v_{j}}{r^{9}}.$$
The last three terms in (\ref{result}) can then be expressed as
\begin{eqnarray*}
\, &\,& \frac{15}{128} \int d\tau  \, v_{k}v_{l} \, (\bar{\psi}_{3}
\Gamma_{0ik}\psi_{3}) \,(\bar{\psi}_{3}
\Gamma_{0jl}\psi_{3})\, \left(63 \frac{x_{i}x_{j}}{r^{11}} - 7
\frac{\delta_{ij}}{r^{9}} \right) \\
= &\,& \frac{15}{128} \int d\tau  \, v_{k}v_{l} \, (\bar{\psi}_{3}
\Gamma_{0ik}\psi_{3}) \,(\bar{\psi}_{3}
\Gamma_{0jl}\psi_{3})\, \partial_{i}\partial_{j}\, \frac{1}{r^{7}}.
\end{eqnarray*}
Using the identification
$$\frac12 (\bar{\psi}_{3}\Gamma_{0ij}\psi_{3}) = J_{ij}
$$ from the spin-orbit interaction,
this is clearly a  higher order spin-orbit interaction.

This result can be compared with that of Morales et al
\cite{Morales2}, who have calculated  spin dependent effects in
scattering amplitudes for  D-branes using
string theoretic methods. In the notation of the present paper, they
find that the  contribution to the scattering amplitude for two D0-branes
which is
quadratic in the angular momentum of the form\footnote{Comparison of
the normalization of the spin-orbit terms  shows that the quantity
$J_{\mu\nu\rho}$ in the paper of Morales et al \cite{Morales2} is
equivalent to $\frac18 (\bar{\psi}_{3} \Gamma_{\mu \nu
\rho}\psi_{3})$ in the present paper.} (in the Minkowski metric, and
after restoring the trace over gauge indices)
$$\frac{1}{192} V_{1} T_{0}^{2}\,\frac{1}{128} \, \mbox{tr} \,  (2  v^{2}\,
G^{i0\mu}G^{j}_{\,\,0 \mu} - v^{2}G^{imn}G^{j}_{\,\,mn} + 4 v^{k}v^{l}G^{i
\mu}_{\,\,\,\,k}G^{j}_{\,\,\mu l})\, \partial_{i}\partial_{j}\,
G_{9}(r),$$
where $V_{1}$ is the D0-brane volume, $T_{0}$ is the tension (mass) and
$G_{9}(r)$ is the nine-dimensional Green's function. Using one of the
identities (\ref{ident1}),
$$  2  \,
G^{i0\mu}G^{j}_{\,\,0 \mu} - G^{imn}G^{j}_{\,\,mn} = 4
G^{i0\mu}G^{j}_{\,\,0 \mu}.
$$
Noting the  fact that there is an implicit $\delta_{00}$ factor on the
$G_{ik\mu}G_{jl\mu}$ term in the identity (\ref{ident4}) which changes
sign on rotation to the Minkowski metric, the result can be rewritten
$$\frac{1}{48} V_{1} T_{0}^{2}\, v^{k}\, v^{l}\, \frac{1}{128} \,\mbox{tr} \,
 (6 G^{i0}{}_{k}G^{j}_{\,\,0 l}
-\delta^{ij}G^{0\mu}{}_{l}G_{0\mu k}+ 2 \delta^{j}_{\,k} G^{0
\mu}{}_{l}G_{0\mu }{}^{i})\, \partial_{i}\partial_{j}\,
G_{9}(r).$$
The last term is a total $\tau$ derivative and so doesn't contribute
to the effective action. The second term vanishes using the fact that
the Green's function is annihilated by the Laplacian. The remaining
term is a higher order spin-orbit interaction of the form obtained in
the matrix model calculation in this paper.
 Comparison with the
spin-independent term in the potential in \cite{Morales2} shows that
$V_{1}T_{0}^{2} G_{9}(r) = \frac{60}{r^{7}},$ which then produces
agreement between  the matrix model
result and the string theoretic result.

\section{Conclusion}
The $\frac{v^{2}\psi^{4}}{r^{9}}$ term in the effective action
(\ref{spin}) has been calculated using the matrix model, and shown to
be consistent with the corresponding term in the scattering amplitude
for   a pair of D0-branes using string theoretic methods. The
technique employed in this paper is well adapted to the underlying
supersymmetry of the matrix model by recognizing that the
one-loop effective action is a superdeterminant. The procedure for
evaluating the superdeterminant is straightforward (but laborious),
in that it only requires the computation of commutators of
supermatrices. In contrast, the method used in \cite{Kraus} and
\cite{Barrio} to calculate the $\frac{v^{3}\psi^{2}}{r^{8}}$ and
$\frac{\psi^{4}}{r^{11}}$ terms respectively
in the effective action involves many
factors  of the full bosonic propagator which are acted on by
derivatives coming from the fermionic propagators. Each of the
vertices also introduces a $\tau$ integral. Although
tractable for the two point function computed in \cite{Kraus}, and
the static case in \cite{Barrio} (where the derivatives from the
fermionic propagators play no role), it is likely to be
cumbersome for higher order amplitudes such as the one computed in
this paper. The technique used here will also extend to the treatment of
more general
Dp-brane scattering amplitudes, involving the computation of the
one-loop effective action in the presence of a fermionic background
for  10D supersymmetric
Yang-Mills theory  reduced to (p+1)-dimensions.

\section*{Appendix A}

Here, spinor identities used in computing the supertrace of the
matrix $M(s)$ determined by the differential equation (\ref{deM}) will
be established. We use the spinor conventions of \cite{GSW}.
The ten-dimensional gamma matrices obey the
anticommutation relations $\{\Gamma_{\mu},\Gamma_{\nu}\} = -2
\eta_{\mu \nu} {\bf 1}_{32},$ with metric $(-,+,+, \cdots , +).$ The spinor
field $\psi$ of ten-dimensional supersymmetric Yang-Mills theory
satisfies the Weyl constraint $(1 - \Gamma_{11}) \, \psi = 0,$ as well
as the Majorana condition $\bar{\psi} = \psi^T C,$ where $C
\Gamma_{\mu} C^{-1} = - \Gamma_{\mu}^T.$ Although the
gamma matrices in ten-diemsions are 32 dimensional, the Weyl
constraint means that attention can be restricted to the appropriate
 16 dimensional chiral projection. Throughout the paper, greek
 letters $ \mu, \nu, \cdots $  take  the values $0, \cdots , 9$ and
 latin letters $i,j, \cdots$
 take the valuess $1, \cdots, 9.$

With $\Gamma_{\mu_1 \cdots \mu_{n}}$ denoting the totally antisymmetric
product of the gamma matrices $\Gamma_{\mu_1}, \cdots,
\Gamma_{\mu_n},$ the fermion bilinears $(\bar{\psi}\Gamma_{\mu_1 \cdots
\mu_n} \psi)$ in the background fermions $\psi = \psi_3 T_3$ vanish
for $n$ even as a consequence of the Weyl constraint. For $n$ odd,
$(C \Gamma_{\mu_1 \cdots \mu_n})_{\alpha \beta}$ is antisymmetric in
the spinor indices $\alpha$ and $\beta$ in the cases $n=3$ and $n=7$ and
otherwise symmetric, so $\psi^{\alpha} (C \Gamma_{\mu_1
\cdots \mu_n})_{\alpha \beta} \psi^{\beta}$ vanishes except in  these
two cases. Since $(\bar{\psi}\Gamma_{\mu_1 \cdots
\mu_7} \psi)$ is proportional to $\epsilon_{\mu_1 \cdots
\mu_7}{}^{\mu_8\mu_9\mu_{10}}\, (\bar{\psi}\Gamma_{\mu_8 \mu_9 \mu_{10}}
\psi),$ the only independent nonvanishing bilinear in the background
fermions is
$$ (\bar{\psi} \Gamma_{\mu_1 \mu_{2}
\mu_3}\psi) \equiv G_{\mu_1 \mu_2 \mu_3}. $$
 Consequently, the Fierz  identity
\begin{equation}
\psi \, \bar{\psi} = - \, \frac{1}{96}\, (\bar{\psi} \Gamma_{\mu_1
\mu_2 \mu_3}\psi) \,\,  P_{-}\, \Gamma_{\mu_1 \mu_2 \mu_3}\, P_{+}
\label{Fierz}
\end{equation}
applies, where $P_{\pm} = \frac12(1\pm \Gamma_{11}).$  This
 is proved by multiplying both sides by $\Gamma_{\rho_1 \rho_2
\rho_3}$ and tracing, not forgetting that
 Weyl projectors make the Gamma matrices effectively 16 dimensional.

In computing the one-loop effective action, a rotation to the
Euclidean metric is made, and the gamma matrices then obey $\{\Gamma_{\mu},
\Gamma_{\nu}\} = -2
\delta_{\mu \nu} {\bf 1}_{16},$ where Weyl projectors are implicit on
the gamma matrices. The only potentially nonvanishing bilinears are
$(\bar{\psi} \Gamma_{\mu} \Gamma_{\nu} \Gamma_{\rho} \psi),$ for which
the
totally antisymmetric piece is $G_{\mu \nu \rho}$ defined above, and
for which any symmetric piece vanishes.
The contraction of $G_{\mu \nu \rho}$ with an
expression symmetric in a pair of indices (such as $\delta_{\mu \nu}$
and $(F^2)_{\mu \nu}$) is thus zero. The following identities are also true:
\begin{equation}
G_{\mu \nu \rho} \, G_{\mu \nu \rho}  =  0, \,\,\,\,
G_{\rho \mu \nu } \, G_{\sigma \mu \nu }  =  0.
\label{ident1}
\end{equation}
The first follows as a consequence of the second. The second is proved
by contracting the Fierz identity  on the left with
$\bar{\psi} \Gamma_{\rho}\Gamma_{\mu}\Gamma_{\nu}$  and
with $ \Gamma_{\nu}\Gamma_{\mu}\Gamma_{\sigma}$ on the right, and
using the identity \cite{GSW}
\begin{equation}
\Gamma_{\nu} \Gamma_{\mu_1 \cdots
\mu_n} \Gamma_{\nu} = (-1)^{(n+1)} (10 - 2n) \Gamma_{\mu_1 \cdots
\mu_n}
\label{gam}
\end{equation}
twice. By applying the Fierz identity ``in reverse'' on the result,
 one obtains an expression proportional $(\bar{\psi}\Gamma_{\rho}\psi)\,
 (\bar{\psi}\Gamma_{\sigma}\psi),$  which vanishes. This argument only
 works because two gamma matrices are moved through each of the
 projectors in the Fierz identity in order to use the result
 (\ref{gam}); if one attempts to apply the same trick to $G_{\rho
 \sigma \mu}G_{\gamma \delta \mu},$ only one gamma matrix is moved
 through each of the projectors, thus interchanging them and preventing the
 use of the Fierz identity ``in reverse.'' So   $G_{\rho
 \sigma \mu}G_{\gamma \delta \mu}$ is in general nonzero.

Another identity which will be used is
\begin{eqnarray}
0 = v_{k}v_{l}\, A_{i}A_{j}\, (&-& 6 \,G_{0ik}G_{0jl} + \delta_{kl}\,
 G_{0i\mu}G_{0j\mu} + \delta_{ij}\, G_{0k\mu}G_{0l\mu} \nonumber \\
 & + &
G_{ik\mu}G_{jl\mu} - 2 \delta_{jk} \, G_{0i\mu}G_{0l\mu} .
\label{ident4}
\end{eqnarray}
This is proved by applying the Fierz identity  to obtain
$$G_{0ik}G_{0jl} = -\, \frac{1}{96} \, G_{\mu_{1}\mu_{2} \mu_{3}} \,
(\bar{\psi} \Gamma_{0}\Gamma_{i}\Gamma_{k}\Gamma_{\mu_{1}\mu_{2}\mu_{3}}
\Gamma_{0}\Gamma_{j}\Gamma_{l}\psi ),$$
and then ``shuffling'' the matrices $\Gamma_{0},$ $\Gamma_{i}$ and
$\Gamma_{k}$ through $\Gamma_{\mu_{1}\mu_{2}\mu_{3}}.$

With $F_{0k}  = v_{k}$ and $H_{\mu i \nu} = (\bar{\psi} \Gamma_{\mu}
\Gamma_i \tilde{F}
\Gamma_{\nu} \psi ),$ it is also possible to prove by similar tricks
that
\begin{eqnarray}
 (F)^{2}_{\mu \nu} G_{\nu i \rho}G_{\rho j \mu} & = &
 v^{2} G_{0i\mu}G_{0j\mu} + v_{k}v_{l}
G_{il\mu}G_{jk\mu} \nonumber \\
 F_{\mu \nu} G_{\nu i \rho}F_{\rho \sigma}G_{\sigma j \mu} & = &
2 v_{k}v_{l} \, G_{0ik}G_{0jl} \nonumber \\
F_{\mu \nu}G_{\nu i \rho}H_{\rho j \mu} & =& -2 v_{k}v_{l}
G_{ik\mu}G_{jl\mu} - 2 v^{2} G_{0i\mu}G_{0j\mu} \nonumber \\
G_{\mu i \rho}F_{\rho \sigma}H_{\sigma j \mu} & = & 4
v_{j}v_{k}G_{0k\mu}G_{0i\mu} - 2 v_{k}v_{l}G_{il\mu}G_{jk\mu} + 4
v_{j}v_{k}G_{0i\mu}G_{0k\mu}\nonumber \\
H_{\mu i \nu}H_{\nu j \mu}\,A_{i}A_{j} & = & 64\,
v_{k}v_{l}\,G_{0ik}G_{0jl}\, A_{i}A_{j}.
\label{GH}
\end{eqnarray}
Only the last of these is difficult to prove, and involves
``shuffling'' gamma matrices so that $\Gamma_{\mu}$ and $\Gamma_{\nu}$
appear together, then applying the Fierz identity and using
(\ref{gam}) twice.

\section*{Appendix B}

In this Appendix, the supertrace of the matrix $M(s)$ defined by
(\ref{deM}) is computed order by order in its power series expansion
in $s.$ Many terms in the expansion have a vanishing supertrace as a
result of the following identities:
\begin{equation}
\mbox{ str}\, N_n = 0, \,\, \mbox{ str}\, Y_n = 0, \,\, \mbox{ str}\,
Z_n = 0,
\label{identities}
\end{equation}
where $Z_{n} = [\tilde{X},Y_{n}].$
We will prove these before proceeding.

Beginning with $\mbox{ str}\, N_n,$ using the definition (\ref{defN})
and the fact that the supertrace of a commutator of  matrices is zero
as a result of the cyclic property of the supertrace,
the only potential contribution is from the $\tau$ derivative in the
commutator $2 [\tilde{X}, Y_{n-1}].$  Using the fact that $N_n$ and
$Y_{1}$  are
independent of $\tau,$ and that
$ Y_1 = - [\tilde{X}, F.\Lambda + Y]$ has a vanishing supertrace
because $\partial_{\tau}( F.\Lambda + Y)$ is off-diagonal, it follows
from (\ref{defY}) that  $2 \partial_{\tau} \mbox{str}\, Y_{n-1} = 4
v^2\, \mbox{str}\, N_{n-2}.$ Thus the vanishing of  $\mbox{str} \, N_n$
follows by induction if $N_2$ and $N_3$ have vanishing supertraces.
The former is easily computed explicitly as
$$ N_{2}  = \left( \begin{array}{cc} 0 & - G_{\mu  0
\rho}\,\bar{\psi} \Gamma_{\rho}\Gamma_{0} \\
0 & 0 \end{array}\right). $$
This obviously has vanishing supertrace, while from (\ref{defN}) and
(\ref{defY}),
$$ \mbox{str} \,N_{3} = 2\, \partial_{\tau}\,\mbox{str} \, Y_{2} = 2\,
 \partial_{\tau}\,\mbox{str}\, [F.\Lambda + Y, Y_{1}] = 0.$$

The vanishing of $\mbox{str}\,Y_n$ follows from the definitions
(\ref{defY}), the fact that $Y_1$ and $N_n$ both have vanishing
supertrace, and that $N_{n}Y_{1}$ vanishes to order $\psi^{4}.$
 Similarly, $ \mbox{str}\, Z_n = \mbox{str}\, [\tilde{X},
Y_n ],$ as the only potential contribution is $\mbox{str} \,
\partial_{\tau} Y_n =  \partial_{\tau}\, \mbox{str}\, Y_n,$ which
vanishes.

We now proceed with the evaluation of the supertrace of $M(s).$
The right hand side of (\ref{deM}) is expanded as a power series in
$s,$ integrated and then exponentiated.  Consider first the terms
in $M(s)$ up to order $s^3:$
\begin{eqnarray*}
 M(s) & = & {\bf 1} + s G  + \frac{s^2}{12}\left( 6 G^2 - N_2 \right) \\
& + & \frac{s^3}{144} \left(24 G^3 -12 G N_2   + 48 v^2 \tau
Y_1 + 12 Y_1^{\,\,2} - 8 Z_2 + N_3 \right).
\end{eqnarray*}
Here, the notation $G = F.\Lambda + Y$ and
$Z_n = [\tilde{X}, Y_n] $ (so $Z_1 = \frac{1}{2} N_2$ by (\ref{defN}))
has been introduced for convenience.
The terms independent of $\psi$ have already been discussed earlier.
The supertrace of the terms dependent on $\psi$ vanishes. This
follows from the identities (\ref{identities}) as
 well as  the specific
 results that $\mbox{str} \,Y^2,$ $\mbox{str}\, ( F.\Lambda) Y^2 $ and
  $\mbox{str}\,Y_1^{\,\,}$  and $\mbox{str} \, G N_{2}$ all vanish.
The latter two are true because, using the  explicit expression for
$N_{2}$ given earlier and
$$Y_1 = \left( \begin{array}{cc} - \frac{i}{2}  G_{\mu 0 \nu} &
F_{\mu \rho}\, \bar{\psi} \Gamma_{\rho} \Gamma_0 \\ 0 & \frac{i}{2}
\Gamma_{\rho} \psi\, \bar{\psi} \Gamma_{\rho} \Gamma_{0} \end{array}
\right),$$
both $\mbox{str} \, Y_{1}^{\,\,2}$ and $\mbox{str} \, G N_{2}$ are
proportional to $G_{\mu 0 \rho}G_{\rho 0 \mu},$ which vashishes by
(\ref{ident1}). In the case of  $\mbox{str}\, Y^2,$
using the expression (\ref{Y}), this is proportional to $v \tau \,G_{\mu 1
\mu} +
b\, G_{\mu 2 \mu},$ which  vanishes due to the antisymmetry of $G_{\mu
\nu \rho}.$. The term $\mbox{str}\, ( F.\Lambda) Y^2,$
is the only nontrivial one, in that it vanishes due to a cancellation
of two potentially nonzero contributions proportional to $G_{012}.$
These are $-2i F_{\mu \nu} G_{\nu i \mu} A_i = - 4 i\, v b \, G_{012}$ and
$\frac{i}{2} (\bar{\psi} \Gamma_{\mu} \Gamma_i F_{\nu \rho}
\Gamma_{\nu \rho} \Gamma_{\mu} \psi) A_i = 4 i\, v b\, G_{012}.$

At order $s^4,$ after eliminating the terms which vanish via the
identities (\ref{identities}),
\begin{eqnarray*}
\mbox{str}\, M(s) & = & \frac{s^4}{288}\, \mbox{str}\,
\left( 12 G^4 + 96 v^2 \tau G Y_1 - 12 G^2 N_{2} + 2 G N_{3} \right. \\
& -& \left.  16 G Z_2  + 24 G Y_1^{\,\,2}  +  N_{2}^{\,\,2}
 + 12 Y_1 Y_2 \right).
\end{eqnarray*}
The last two  terms don't contribute because $\mbox{str}\,
N_{2}^{\,\,2}$ is of order $\psi^{6},$ and
 $\mbox{str}\, Y_1 Y_2 = \mbox{str}\, Y_1 [G,Y_1]$ is
zero using the cyclic property of the supertrace.
The terms involving $\mbox{str} \, GN_{3}$ and  $\mbox{str} \,
GZ_{2}$ can be combined since $\mbox{str} \, GN_{3}= 2 \,\mbox{str} \,
GZ_{2}$ using (\ref{defN}).
 At this point, it is necessary to explicitly evaluate
supertraces. To order $\psi^2,$ only the first two terms contribute,
yielding the bilinears
$(F^2)_{\mu \nu} G_{\nu i \mu}$ and   $(\bar{\psi}
\Gamma_i  \psi),$ both of which vanish by the results of Appendix A.
To obtain the $\psi^4$ terms, an explicit expression for $N_{3}$
must be computed:
$$ N_{3} = \left( \begin{array}{ll} 2 G_{\mu 0 \rho} G_{\rho  0 \nu} &
\,\,\,\,\,4i F_{\mu \rho} G_{\rho 0 \sigma} \bar{\psi}
\Gamma_{\sigma}\Gamma_{0}
- 2 i v G_{\mu 0 \rho}  \bar{\psi} \Gamma_{\rho}\Gamma_{1} \\
0 & \,\,\,\,\,-2 \Gamma_{\rho}\psi G_{\rho 0 \sigma} \bar{\psi}
\Gamma_{\sigma}\Gamma_{0} \end{array} \right).$$
The order $\psi^{4}$ terms are proportional to $G_{i \mu
\nu} G_{j \mu \nu}$ and $G_{00\mu} G_{01 \mu},$ which vanish via
(\ref{ident1}) and the antisymmetry of $G_{\mu \nu \rho}$ respectively.

At order $s^5,$ again eliminating the terms which vanish via the
identities (\ref{identities}), as well as terms of the form $N_n
N_m,$  $N_n Y_m$ and $N_n Z_m$ (which are all of order $\psi^5$),
\begin{eqnarray*}
\mbox{str}\, M(s) & = & s^5 \,\mbox{str} \, \left( \frac{G^5}{120} -
\frac{G^{3}N_{2}}{72}
 + \frac{G^2}{288} ( N_{3} + 48 v^2 \tau Y_1  +
12 Y_1^{\,\,2} - 8 Z_2 ) \right. \\
& + & \frac{G}{2880} ( 160 v^{2} N_{2} + 3 N_4 + 240 v^2 \tau Y_2 +
60 Y_1 Y_2  + 60 Y_2 Y_1 - 30 Z_3 ) \\
& + & \left. \frac{1}{7200} (-480 v^2 Y_1^{\,\,2} + 40 Y_2^{\,\,2} +60
Y_1 Y_3 ) \right).
\end{eqnarray*}
This expression can be simplified somewhat. Using $Y_2 = [G,Y_1],$ it
follows that $\mbox{str} \, G Y_2$ vanishes by the cyclic property of
the supertrace. Similarly, $\mbox{str} \, G N_4 = 2 \,\mbox{str} \, G Z_3$
using (\ref{defN}). Using (\ref{defY}) and the Jacobi identity, this
can be further simplified to $\mbox{str} \, G N_4 = \mbox{str} \,G (
8 v^{2}N_{2} + 2
 [Y_2,Y_1]).$ By a similar procedure, $\mbox{str} \, G^2 N_3 = 2\,
\mbox{str}\,
G^2 [Y_1,Y_1] = 0.$ Also $\mbox{str} \, Y_1 Y_3 = 4 v^2 \,\mbox{str} \,
Y_1^2 - \mbox{str} \,G [Y_1,Y_2].$
To order $\psi^2,$ the only potential contributions are from
$\frac16 v^2 \tau\, \mbox{str}\, G^2 Y_1 + \frac{1}{120}\, \mbox{str}\, G^5,$
yielding
$$ \frac{1}{120}\, \mbox{tr} \, \left( 40 i (F)^{3}_{\mu \nu}G_{\nu i
\mu} - 10 i (F)^{2}_{\mu \nu} H_{\nu i \mu} - 10i v^{2}F_{\mu
\nu}G_{\nu i \mu} + \frac{5i}{2} v^{2} H_{\mu i \mu} \right) A_{i},$$
where the trace is over gauge indices and
$$ H_{\mu i \nu} = (\bar{\psi} \Gamma_{\mu} \Gamma_i \tilde{F}
\Gamma_{\nu} \psi ).$$
At order $\psi^4,$ there is a nontrivial cancellation between
potentially nonzero terms proportional to $v^{2}G_{01\mu}G_{01\mu}$
to give a net result of zero.

Having identified the leading $\psi^2$ term at order $s^5$ in the
expansion of $M(s),$ only $\psi^4$ terms need be considered at order
$s^6.$ Again eliminating terms which are obviously zero,
\begin{eqnarray*}
\mbox{str}\, M(s) & = & s^6 \,\mbox{str} \, \left( \frac{G^6}{720}
- \frac{G^{4}N_{2}}{288}  + \frac{1}{864} G^3 (N_{3} +  48 v^2 \tau Y_1 + 12
Y_1^2 - 8Z_2) \right. \\
& + & \frac{1}{5760} G^2 ( 160 v^{2}N_{2}+  3 N_4 + 240 v^2 \tau Y_2 +
120 Y_1 Y_2 - 30 Z_3) \\
& +& \frac{1}{7200} G ( -480 v^{4}\tau^{2}N_{2} + 20 v^{2}N_{3}+  N_5 - 1440
v^4 \tau Y_1 - 480 v^2 Y_1^{\,\,2}\\
& +& 40 Y_2^{\,\,2} + 120 v^2 \tau
Y_3 + 60 Y_1 Y_3 + 160 v^2 Z_2 - 12 Z_4) \\
&+&\left. \frac{1}{2160} ( -80 v^2 Y_1 Y_2 + 5 Y_2 Y_3 + 3 Y_1 Y_4
+ 120 v^4 \tau^2 Y_1^{\,\,2}) \right).
\end{eqnarray*}
Many of the terms can be simplified using the definitions
(\ref{defN}) and (\ref{defY}) together with the Jacobi identity and
the fact that $\mbox{str}\, G^n [W,G] = 0$ for any supermatrix $W.$
One finds that $\mbox{str}\,G^2 Y_2,$ $\mbox{str}\,G^3 Z_2,$
$\mbox{str}\, G^{3}N_{3},$
$\mbox{str}\, Y_1 Y_2,$ $\mbox{str} \,Y_2 Y_3,$ and $ \mbox{str}\, Y_1
Y_4$ all vanish. Also,
\begin{eqnarray*}& &\mbox{str}\, G(N_5 - 12Z_4) = -20\, \mbox{str} \, G
N_3 = 0,\\
 & & \mbox{str}\, G Y_3 = 2 v^{2}\tau \, \mbox{str} \, GN_{2} + 4 v^2 \,
\mbox{str} \, G Y_1,\\
 & &\mbox{str} \, G^2(3N_4 - 30 Z_3) = -96 v^{2}\mbox{str}\, G^{2}N_{2}
 - 24 \,\mbox{str} \, G^2 [Y_2,Y_1].
\end{eqnarray*}
 On computing the remaining
supertraces, one finds a nonvanishing contribution only from $\frac{1}{720}
\,\mbox{str}\, G^6,$ which yields
\begin{eqnarray*}
\frac{1}{720} &\,& \mbox{tr}\, \left( - 12 F_{\mu \nu} G_{\nu i \rho}
F_{\rho \sigma} G_{\sigma j \mu} - 24 (F^{2})_{\mu \nu}G_{\nu i \rho}
G_{\rho j \mu} \right. \\
& + &  \left. 6 F_{\mu \nu} G_{\nu i \rho} H_{\rho j \mu} + 6  G_{\mu i \nu}
F_{\nu \rho} H_{\rho j \mu}
 - \frac{3}{4} H_{\mu i \nu} H_{\nu
j \mu} \right) \, A_i A_j ,
\end{eqnarray*}
with $H_{\mu i \nu} = (\bar{\psi} \Gamma_{\mu} \Gamma_i \tilde{F}
\Gamma_{\nu} \psi ).$

\vspace{2cm}

{\bf Note Added:} hep-th/9806081 has just appeared, which extends the work
of  \cite{Morales2} and compares  it with 11-dimensional supergravity.

\end{document}